\documentclass{article}

\usepackage[final, nonatbib]{nips_2017}

\usepackage[utf8]{inputenc} % allow utf-8 input
\usepackage[T1]{fontenc}    % use 8-bit T1 fonts
\usepackage{hyperref}       % hyperlinks
\usepackage{url}            % simple URL typesetting
\usepackage{booktabs}       % professional-quality tables
\usepackage{amsfonts}       % blackboard math symbols
\usepackage{nicefrac}       % compact symbols for 1/2, etc.
\usepackage{microtype}      % microtypography

\usepackage{graphics} % for pdf, bitmapped graphics files
\usepackage{amsmath} % assumes amsmath package installed
\usepackage{amssymb}  % assumes amsmath package installed
\usepackage{color}
\usepackage{wrapfig}
\usepackage{mathtools}
\usepackage{float}

\title{Exploiting Nontrivial Connectivity for Automatic Speech Recognition}

\author{ 
Marius Paraschiv\\
Corti, Copenhagen, Denmark\\
\texttt{mp@cortilabs.com} 
\And
Lasse Borgholt\\
Corti, Copenhagen, Denmark\\
\texttt{lb@cortilabs.com}
\And
Tycho Max Sylvester Tax\\
Corti, Copenhagen, Denmark\\
\texttt{tt@cortilabs.com}
\And
Marco Singh\\
Corti, Copenhagen, Denmark\\
\texttt{ms@cortilabs.com}
\And
Lars Maal{\o}e\\
Technical University of Denmark\\
Corti, Copenhagen, Denmark\\
\texttt{lm@cortilabs.com}
}

\begin{document}

\maketitle
\thispagestyle{empty}

\setcounter{page}{1}
\pagenumbering{arabic}

%%%%%%%%%%%%%%%%%%%%%%%%%%%%%%%%%%%%%%%%%%%%%%%%%%%%%%%%%%%%%%%%%%%%%%%%%%%%%%%%
\begin{abstract}

Nontrivial connectivity has allowed the training of very deep networks by addressing the problem of vanishing gradients and offering a more efficient method of reusing parameters. In this paper we make a comparison between residual networks, densely-connected networks and highway networks on an image classification task. Next, we show that these methodologies can easily be deployed into automatic speech recognition and provide significant improvements to existing models.
\end{abstract}

%%%%%%%%%%%%%%%%%%%%%%%%%%%%%%%%%%%%%%%%%%%%%%%%%%%%%%%%%%%%%%%%%%%%%%%%%%%%%%%%
\section{Introduction}

Deep neural networks have become the standard in end-to-end speech recognition systems. Models such as DeepSpeech 2 \cite{ds2} have replaced cumbersome hand-engineered pipelines while being much more flexible in terms of deployment in practical applications. A more recent proposal is Wav2Letter \cite{w2l,w2l_tl}, a deep fully-convolutional network, providing near state-of-the-art results on speech recognition tasks. 

Empirically, it was shown that deeper convolutional networks tend to offer improved performance. One of the earliest convolutional networks, LeNet5 \cite{leNET}, consisted of only 5 layers. Since then, deeper architectures, such as the VGG \cite{VGG} architecture with 19 layers, have shown improved performance on image classification tasks. More recently, Highway Networks \cite{highwayNET1,highwayNET2} and Residual Networks \cite{resnet}, reaching over 1000 layers, continued to improve classification accuracy. One of the latest propositions in the family of very deep neural networks is the Densely Connected Convolutional Network \cite{densenet,densenet2}. The last three models, from now on denoted as HighwayNETs, ResNETs and DenseNETs, respectively, are of particular interest. 

A significant difference between the above three architectures and regular neural networks is their use of skip-connections (or nontrivial connections). ResNETs use identity connections that skip a block of layers, passing the input of a previous nonconsecutive layer to the current one. HighwayNETs are very similar, but make use of gated connections, an idea reminiscent of LSTMs \cite{lstm} while DenseNETs connect every layer to the outputs of every previous layer, using identity connections. Given the performance of the above architectures on image recognition problems and the increasing use of fully-convolutional networks in ASR tasks, studying the effects of nontrivial connectivity on convolutional networks in the context of speech recognition is justified.

According to the Universal Approximation theorem \cite{approx1,approx2}, a single-layer network with a large (but finite) number of units can act as an approximator for highly complex functions. By reducing layer width and increasing the overall number of layers, the network gains the ability to learn increasingly abstract features. It has been proven both experimentally \cite{resnet,highwayNET1,densenet} and theoretically \cite{power_depth} that increased depth comes with improved performance. 

Training deep networks brings forth a set of challenges, most notably the vanishing gradient problem. As a network increases in depth, it can happen that the gradients of the network's output with respect to the parameters of initial layers become very small. This is an obstacle that prevents efficient training of extremely deep networks. Adding identity or gated skip-connections provides shortcuts for gradient backpropagation, which facilitate network training. A second advantage to adding skip-connectivity is that this allows the network to better reuse features. Thus, we are also interested in net gains due to the more complex architecture exclusively while maintaining depth constant. 

We start by training all three models with 56 and 100 layers on the CIFAR-10 dataset. The reason for this is that no direct comparison between these three architectures can be found in the literature. By maintaining the core of the network and all hyperparameters identical, and varying only the type of skip-connections, we can study the impact of these connections on the overall results. Then we show that similar performance gains can be achieved when training the networks for ASR, even without increasing network depth. While this comparison is far from proving the superiority of one architecture over the others, it does show that, by using more complex connectivity, gains can be achieved on two different tasks.

%%%%%%%%%%%%%%%%%%%%%%%%%%%%%%%%%%%%%%%%%%%%%%%%%%%%%%%%%%%%%%%%%%%%%%%%%%%%%%%%
\section{Method}

This section provides a brief overview of the selected architectures and of some notable implementation particularities. 

\subsection{Residual Networks}

Residual Networks \cite{resnet} represent a slight variation from a traditional neural network due to their use of identity skip-connections. In this paper, skip-connections for both ResNETs and HighwayNETs jump over 2 layers.

The skip-connections of residual networks can be understood mathematically as a reformulation of the mapping to be learned. If the desired underlying mapping is $\text{G}(\text{x})$, representing the function one wishes to approximate, the skip-connections allow the residual block to learn the mapping $\text{G}(\text{x}) := \text{F}(\text{x}) + \text{x}$ instead, in terms of a residual function $\text{F}(\text{x})$, representing the activation of the block layers and $\text{x}$ representing the unchanged output of the previous network block. 

\subsection{Highway Networks}

Another architecture that uses skip-connections is the HighwayNET \cite{highwayNET1}. The main difference between it and the ResNET is that a HighwayNET uses gated connections. 

A regular network consists of a series of layers, each of which applies a non-linear transformation $\textnormal{H}(\textnormal{x}, \textnormal{W}_{\textnormal{H}})$ to the activation of the previous layer, $\text{y} = \text{H}(\text{x}, \text{W}_{\text{H}})$, where $\text{W}_{\text{H}}$ are the weights parameterizing the non-linear function H, $\text{x}$ is the output of the previous layer and $\text{y}$ is the current layer activation. A HighwayNET contains two additional transformations, $\text{T}(\text{x}, \text{W}_{\text{T}})$ and $\text{C}(\text{x}, \text{W}_{\text{C}})$, which give rise to a composition rule : $\text{y} = \text{H}(\text{x}, \text{W}_{\text{H}}) \cdot \text{T}(\text{x}, \text{W}_{\text{T}}) + \text{x} \cdot \text{C}(\text{x}, \text{W}_{\text{C}})$.

The additional transformations $\text{T}$ and $\text{C}$ are known as the \textit{transform gate} and the \textit{carry gate}, respectively, and $\text{W}_{\text{T}}$ and $\text{W}_{\text{C}}$ represent their corresponding weights. The role of the transform gate T, is to learn how many of the features of the previous layer need to be passed through the nonlinearity of the current layer, and the role of the carry gate C is to determine how many of these features should remain unchanged. This extra degree of freedom provided to the network comes at the cost of additional parameters. In practice, this translates to slower convergence, compared to the ResNET. A method for reducing additional parameters is not to use a carry gate altogether, and to transform the above equation into, $\text{y} = \text{H}(\text{x}, \text{W}_{\text{H}}) \cdot \text{T}(\text{x}, \text{W}_{\text{T}}) + \text{x} \cdot (1 - \text{T}(\text{x}, \text{W}_{\text{T}}))$.

In order to ensure that the network will use the alternate gate connection, one must initialize it with a negative bias, proportional to the depth of the network. Written in detail, the transform gate is of the form, $\text{T}(\text{x}) = \sigma(\text{W}_\text{T} ^\top \cdot \text{x} + \text{b}_\text{T})$, where $\text{b}_\text{T}$ is the gate bias, $\sigma(\text{x})$ is the sigmoid function and $\text{W}_\text{T} ^\top$ is the transpose of the weight matrix, corresponding to the transform gate. As such, the product between the gate $\text{T}(\text{x})$ and $\text{H}(\text{x})$ is a product between a scalar and a tensor.  The bias term must be initialized with a negative value, depending on network depth, such that the network is biased towards carry behavior. If a negative bias is not used, the network will tend to ignore the gated connections. In both residual and highway networks, we have used skip-connections over blocks of two layers. We have also experimented with blocks containing bottleneck layers, but with no noticeable improvements. 

\subsection{Densely Connected Networks}

An extreme case of using skip-connections is the DenseNET \cite{densenet}. Instead of having skip-connections over 2 layers, the network is divided into densely-connected blocks. Every layer in a block is connected to every previous layer. 

\begin{figure}
\begin{minipage}[t]{0.45\hsize}\centering
\begin{center}
        \includegraphics[width=\textwidth]{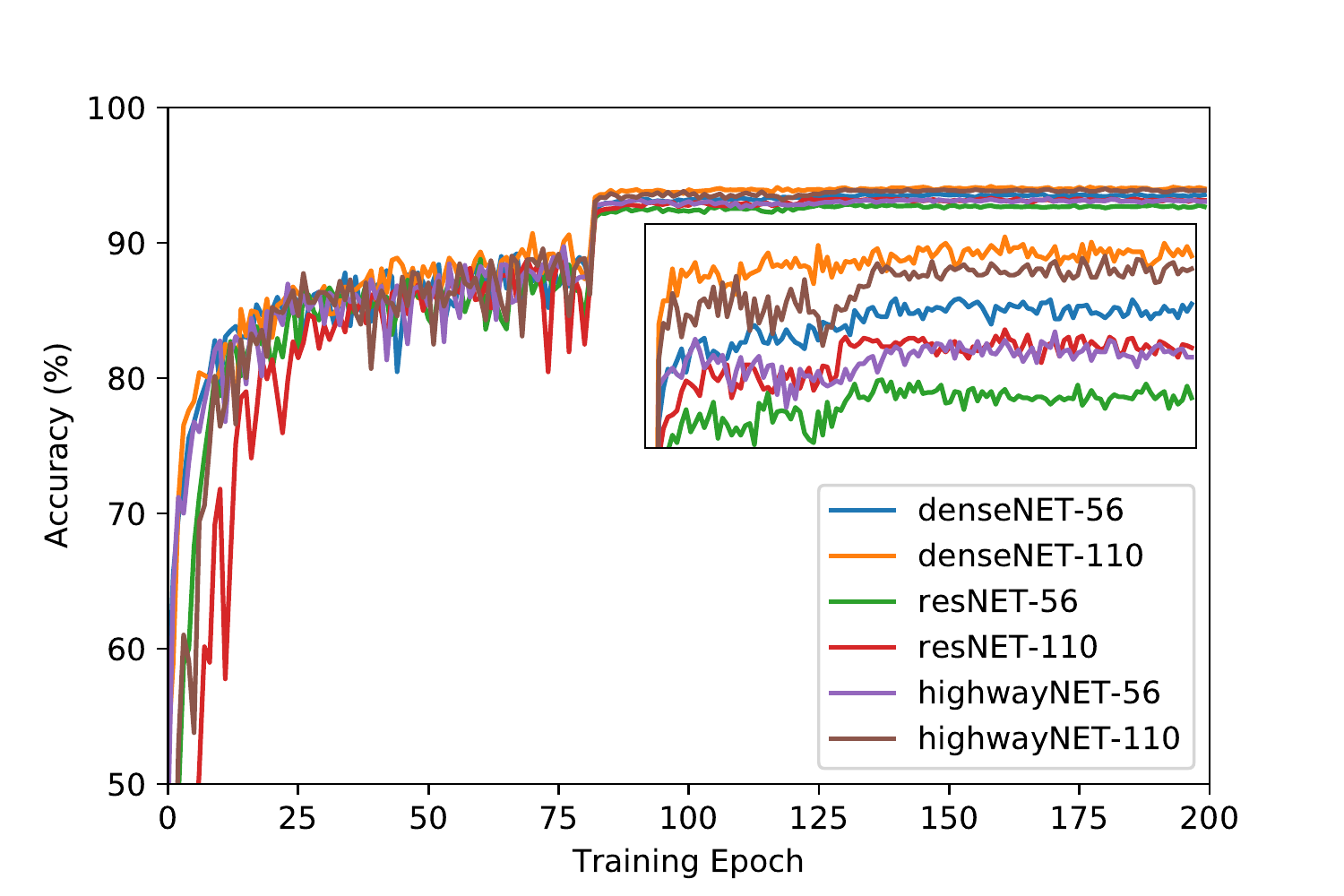}
\end{center}
\caption{Classification accuracy in terms of the number of training epochs, for the CIFAR-10 dataset. At epochs 82 and 123, the learning rate is decreased by a factor of 10, and the enlarged plot section shows relative network performance, starting from epoch 82.}
\label{cifar10loss}
\end{minipage}
    \hfill
\begin{minipage}[t]{0.45\hsize}\centering
\begin{center}
        \includegraphics[width=\textwidth]{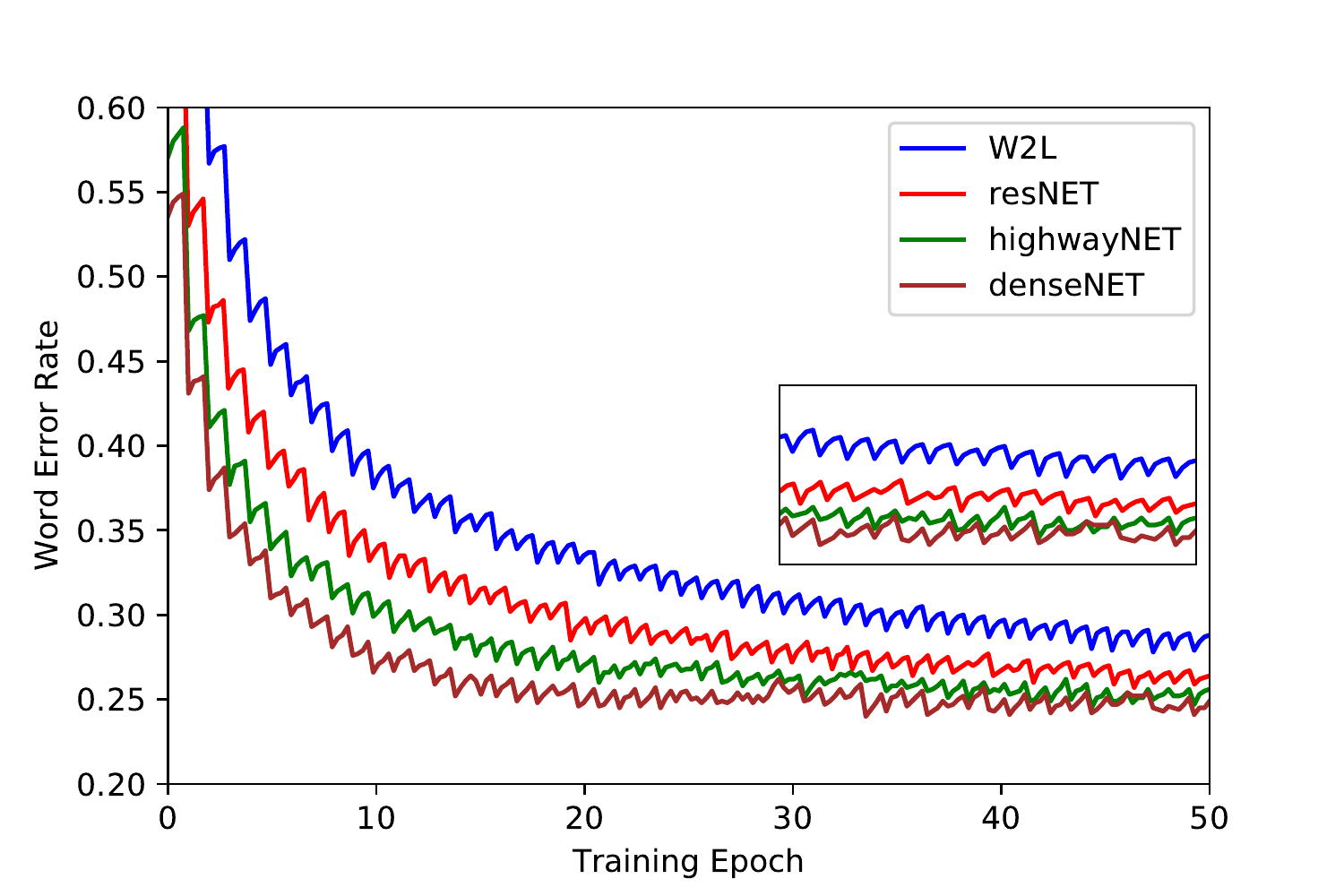}
\end{center}
\caption{Validation word error rates for the baseline Wav2Letter model and the three derived networks, in the absence of a language model, for the first 50 training epochs.}
\label{wav2lett}
\end{minipage}
\end{figure}

Current activations are then concatenated to previous ones and passed on to the next layer. In this way, a layer is able to append newly learned features to already existing ones instead of modifying them. One can define DenseNET activations in terms of a composite function depending on the overall concatenated tensor of features, $\text{y}_\text{$\ell$} = \text{H}[\text{x}_0, \text{x}_1, ... ,\text{x}_{\text{$\ell$}-1}]$, where $\text{y}_\text{$\ell$}$ is the activation of layer $\text{$\ell$}$.

Batch normalization applies scaling and bias to the input features, and by placing the convolution last, preceded by the nonlinearity and the batch normalization, every layer is free to apply a different scaling to the same features. This has been shown in Ref.~\cite{densenet2} to have a considerable impact on the network's classification accuracy.

Due to the concatenation operation, performed after every layer, the concatenated feature vector has the tendency to grow very large, and performing convolutions with larger kernels quickly becomes impractical. For this reason, the network is split into densely-connected blocks, and before passing the concatenated features to the next block, we use a dimensionality reduction layer (also called a transfer layer) consisting of 1D convolutions with kernel size of 1. 

\section{Image Classification}

% Image classification results

\begin{table}[ht]
\begin{minipage}[b]{0.45\hsize}\centering
  \caption{Network classification accuracy on the CIFAR-10 dataset.}
  \label{image_results}
  \begin{tabular}{ll}
    \toprule
    Architecture     & Accuracy (\%)   \\
    \midrule
    ResNET(56)      & $92.80 \pm 0.13$      \\
    HighwayNET(56)  & $93.29 \pm 0.08$      \\
    DenseNET(56)    & $93.37 \pm 0.13$      \\
    \midrule
    ResNET(110)      & $93.35 \pm 0.19$      \\
    HighwayNET(110)  & $93.96 \pm 0.17$      \\
    DenseNET(110)    & $94.04 \pm 0.11$      \\    
    \bottomrule
  \end{tabular}
    \end{minipage}
    \hfill
    \begin{minipage}[b]{0.45\hsize}\centering
     \caption{Word error rates for the baseline and non-trivially-connected models.}
     \label{wer_results}
    \begin{tabular}{ll}
    \toprule
    Architecture     & WER  \\
    \midrule
    W2L (our benchmark) & $18.6$      \\
    ResNET-W2L      & $17.2$      \\
    HighwayNET-W2L  & $14.5$      \\
    DenseNET-W2L    & $13.7$      \\
    \bottomrule
  \end{tabular}
    \end{minipage}
\end{table}

For the task of image classification we use the CIFAR-10 dataset. For simplicity, the architectures are split into blocks of 2 convolutional layers each. The initial layer and first $n$ blocks are identical, comprised of layers with 16 filters each. These are followed by two groups of $n$ convolutional blocks with 32 and 64 filters, respectively.

Striding is performed in the first layer in the second and third group of convolutional blocks. We have experimented with $n=9$ and $n=18$, corresponding to 56 and 110-layer networks, respectively. The sequential blocks are followed by global average pooling and a softmax layer.

We present the test classification accuracies of HighwayNET, ResNET and DenseNET models, in terms of training epoch, in Fig.\ref{cifar10loss}. The training procedure is the same as in Ref.~\cite{resnet}, for all models, and results are presented in Table~\ref{image_results}.

\section{Speech Recognition}

For automatic speech recognition, the LibriSpeech dataset, consisting of audio-book fragments from the public domain and their corresponding transcriptions, has been used, with a partition of 360 hours of clean speech. We start with Wav2Letter \cite{w2l} as our baseline model, trained on spectrogram inputs. It is a fully-convolutional architecture, consisting exclusively of 1-D convolutions in the time dimension. 

Striding is performed by the first convolutional layer, followed by 7 identical layers with relatively small kernels. Onto these layers identity skip-connections, corresponding to ResNETs and DenseNETs, or gated skip-connections, corresponding to HighwayNETs, are added. This exact structure is preserved for all models, thus clearly showing the contributions brought by the extra connections. Finally, the features are passed through a large-kernel layer and a convolutional layer with a kernel size of 1.

We use the CTC loss function \cite{ctc} and network output is decoded with the beam search procedure \cite{beam_search} utilizing a modified Kneser-Ney smoothed 4-gram model \cite{kenlm} trained on the original training data and the additional LibriSpeech language modeling data.

Table. \ref{wer_results} contains the word error rates for the baseline model and the three selected architectures on the 360-hour dataset. Our benchmark results for the baseline model are slightly higher than the WER of 14 that the original paper \cite{w2l} reported, due to the fact that training method and hyperparameters used were not specified. As initially asserted, the performance gains through the use of nontrivial connections are task-independent as are the relative performance differences between the three models. A comparative plot of validation WER during training, for the four models is given in Fig.\ref{wav2lett}.

Our results show that all three networks with skip-connectivity outperform the baseline W2L model, even while maintaining the same depth. Feature reuse plays an important role. ResNETs and HighwayNETs aggregate features of different layers by addition while the DenseNET  concatenates all previous features, hence giving one layer access to the outputs of all previous ones in the block.

Continuing with our investigation, we trained the same models on the full LibriSpeech 1000-hour clean speech data, and our baseline model achieved a word error rate of \textbf{9.1} compared to the 9.4 reported in the original paper \cite{w2l}. Due to time constraints, the results obtained by the models on this dataset, as well as the exploration of very deep networks, will be the subject of a future publication.

\section{Conclusion}

By adding skip-connectivity and improving feature reuse, a consistent improvement in network performance is seen, even without increasing network depth. The next step is to explore both the depth and width dimensions \cite{wide_resnet}, in the context of speech recognition, and observe how such architectures are aided by more complex connections.

%%%%%%%%%%%%%%%%%%%%%%%%%%%%%%%%%%%%%%%%%%%%%%%%%%%%%%%%%%%%%%%%%%%%%%%%%%%%%%%%

\end{document}